\documentclass[]{raa}            
\usepackage{graphicx,times}
\usepackage{natbib}

\begin{document}

\def\be{\begin{equation}}
\def\ee{\end{equation}}
\def\bq{\begin{eqnarray}}
\def\eq{\end{eqnarray}}
%........................................................

\title{Mysteries of the geometrization of gravitation
\footnote{Expanded version (with new findings{ added})
of the essay (arXiv:1206.2795) awarded `Honorable Mention'
 of the year 2012 by the Gravity Research Foundation.}}

 \volnopage{ {\bf 2013} Vol. {\bf 13} No. {\bf 12}, 1409--1422}
\setcounter{page}{1409}

\author{Ram Gopal Vishwakarma}

  \institute{Unidad Acad$\acute{e}$mica de Matem$\acute{a}$ticas,
 Universidad Aut$\acute{o}$noma de Zacatecas,
 C.P. 98068, Zacatecas, ZAC,
 Mexico; {\it vishwa@uaz.edu.mx}\\
\vs\no {\small Received 2012 September 22; accepted 2013 June 28} }

\abstract{   As we now know, there are at least two major
difficulties with {g}eneral {r}elativity (GR). The first one is
related {to} its incompatibility with quantum mechanics, in the
absence of a{ consistent,} widely accepted theory that combines
the two theories. The second problem is related {to} the
requirement of the dark sectors$-$inflaton, dark matter and dark
energy by the energy-stress tensor, which are needed to explain a
variety of astronomical and cosmological  observations. {Research
has indicated that }the dark sectors themselves do not have any
non-gravitational or laboratory evidence{. Moreover,} the dark
energy poses, in addition, a serious confrontation between
fundamental physics and cosmology. Guided by theoretical and
observational evidences, we are led to a{n} {idea that} the source
of gravitation and its manifestation in GR{ should be modified}.
The result is in striking agreement with not only the theory, but
also the observations, without requiring the dark sectors of the
standard approach. Additionally, it provides natural explanations
to some unexplained puzzles. }

\authorrunning{R. G. Vishwakarma}

 \titlerunning{Mysteries of the Geometrization of Gravitation}

\maketitle

\section{Introduction}

Einstein's theory of general relativity (GR), which provides the
current description of gravitation in modern physics, ranks as one
of the crowning intellectual achievements of the twentieth
century. It is a geometric theory of gravitation which describes
gravity not as a `force' in the usual sense but as a manifestation
of the curvature of spacetime. In particular, the curvature of
spacetime is directly related to the energy-stress tensor
$T^{\mu\nu}$ through the Einstein field equations{ defined by}
\be
R^{\mu\nu}-\frac{1}{2}R~ g^{\mu\nu}=-\frac{8\pi G}{c^4}T^{\mu\nu},
~~~~~(\mu,\nu=0,1,2,3)\, .\label{eq:EinsteinEq}
\ee
 It should be
noted that Equation~(\ref{eq:EinsteinEq}) attributes the source of
curvature entirely to matter, as the tensor $T^{\mu\nu}$ does not
include the energy, momenta {or} stresses associated with the
gravitational field itself (since a proper energy-stress tensor of
the gravitational field does not exist), though it does
incorporate all the candidates of material fields including dark
energy.

Although GR is not the only relativistic theory of gravitation, it
is the simplest theory that has survived the tests of nearly a
century of observational confirmation ranging from  the solar system
to the largest scales,{ including} the universe itself. However, this success is
achieved provided we admit three completely independent new
components in the energy-stress tensor $-$  inflaton, dark matter
and dark energy, which are believed to play significant roles in the
dynamics of the universe during their turns.  However, there {has been},
until now, no non-gravitational or laboratory evidence for any of
these dark sectors. Additionally, the mysterious dark energy, which
has been {evoked} primarily to fit the observations of{ type Ia} supernovae
(SNeIa), poses a serious confrontation between
fundamental physics and cosmology.

Despite the remarkable success of GR, many researchers interpret
the observations supporting the requirement of the dark sectors as
a failure of the theory. This reminds us of Einstein's `biggest
blunder' when he forced his equations to predict the unstable
static universe (by imposing the cosmological constant) though a
more natural implication of GR -- the expanding universe -- was
already known to him. It appears that we have encountered a
similar situation when we are trying to explain the observations
in terms of the dark sectors. Though there have been other simpler
explanations, they have not been paid proper attention. For
example, it has been known since the first generation of SNeIa
observations that the data are consistent with the `vacuum'
{Friedmann-Robertson-Walker} model ($\Omega_{\rm m}=0=\Lambda$).

While Einstein's blunder was perhaps motivated by his religious
conviction that the universe must be eternal and unchanging, in the
present case it is one's deep-rooted conviction that space would
remain empty unless it is filled with $T^{\mu\nu}$. But what are the
reasons to doubt this obvious and well-established notion? We shall
see in the following that{ expressions derived from} Equation~(\ref{eq:EinsteinEq}) do not
necessarily represent an empty space in the absence of $T^{\mu\nu}$
and the sources of gravitation do exist there in the form of the
geometry itself. Although this may appear orthogonal to the usual
understanding, it is strongly supported by observations,
ranging from the solar system to the largest scales, which seem to
favor Equation~(\ref{eq:EinsteinEq}) without $T^{\mu\nu}$,
implying that the tensor is not needed{.}

Then, let us first see {how} the `vacuum' field equations{ are
supported by observations}. {W}hat are the observations/experiments
which have directly tested {the complete Einstein's
}Equation~(\ref{eq:EinsteinEq})? The classical tests of GR consider
$T^{\mu\nu}=0$. The same is true for the more precise tests of GR
made through the observations of radio pulsars, which are rapidly
rotating strongly magnetized neutron stars. The pulsar tests assume
the neutron stars {to be} point-like objects and look for the
relativistic corrections in the post-Keplerian parameters by
measuring the pulsar timing. The tests do not even require
know{ledge of} the exact nature of the matter that pulsars and other
neutron stars are made of. As $T^{\mu\nu}=0$ implies $T=0=R$ in
which case Equation~(\ref{eq:EinsteinEq}) reduces to \be
R^{\mu\nu}=0,\label{eq:RicciEq} \ee all we can claim is that it is
only Equation~(\ref{eq:RicciEq}) which has been verified by the
classical tests of GR. As these tests have{ only} been limited to
our galaxy, let us see how this equation fairs against the
cosmological observations. For this purpose, let us first solve
Equation~(\ref{eq:RicciEq}) for a homogeneous and isotropic
spacetime, as is expected  on a large enough scale.  Obviously, the
considered symmetry of homogeneity and isotropy requires the metric
to be the Robertson-Walker one given by
\be ds^2=c^2
dt^2-S^2\left(\frac{dr^2}{1-kr^2}+r^2d\theta^2+r^2\sin^2\theta
~d\phi^2\right)\, ,\label{eq:RW}
\ee
 where $S(t)$ is the scale
factor of the universe. For the metric (\ref{eq:RW}), {expressions
derived from }Equation~(\ref{eq:RicciEq}) yield
\begin{eqnarray}
R^0_{~0}&=&\frac{3}{c^2}\frac{\ddot{S}}{S}=0\, , \\
%\ee \be
R^1_{~1}&=&R^2_{~2}=R^3_{~3}=\frac{1}{c^2}
\left(\frac{\ddot{S}}{S}+2\frac{\dot{S}^2}{S^2}+2kc^2\frac{1}{S^2}\right)=0\,,
%\ee
\end{eqnarray}
 which uniquely determine \be S=ct ~~{\rm with} ~~
k=-1,\label{eq:scale} \ee so that the final solution reduces to
\be ds^2=c^2
dt^2-c^2t^2\left(\frac{dr^2}{1+r^2}+r^2d\theta^2+r^2\sin^2\theta
~d\phi^2\right).\label{eq:milne} \ee

\section{Support from the Cosmological Observations to $R^{\mu\nu}=0$}

\subsection{Observations of SNeIa}

In order to study the compatibility of Equation~(\ref{eq:milne})
with the cosmological observations, let us first consider the
observations of SNeIa. {The s}olution{ given in Equation} (\ref{eq:scale}) is efficient enough
to define uniquely, without requiring any inputs from the matter
fields, the luminosity distance $d_{\rm L}$ of a source {with} redshift
$z$ by \be d_{\rm L}=cH_0^{-1}(1+z)\sinh[\ln(1+z)],\label{eq:d_L}
\ee where $H_0$ represents  the present value of the Hubble
parameter $H=\dot{S}/S$. It is already known that this model, albeit
non-accelerating ({and not} decelerating), is consistent with the
observations of SNeIa {\it without requiring any dark energy}. As
early as 1998, the Supernova Cosmology Project team noticed from
the analysis of their first-generation SNeIa data
that the performance of the empty model ($\Omega_{\rm
m}=0=\Omega_\Lambda$) is practically identical to that of the
best-fit unconstrained cosmology with a positive $\Lambda$
\citep{perlmutter}.
 Let us consider a newer dataset\footnote{Although various newer SNeIa datasets are available, however, the way they are analyzed has left little scope for testing a theoretical model with them. This issue has been addressed in \cite{critique}.}, for example, the  `new gold sample' of 182 SNeIa \citep{riess}, which is a reliable set of SNeIa with reduced calibration errors arising from the systematics.
The model (\ref{eq:d_L}) provides an excellent fit to the data
with a value of $\chi^2$ per degree of freedom (DoF)
$=174.29/181=0.96$ and a probability of the goodness of fit
$Q=63\%$. Obviously the standard $\Lambda$CDM model has {an }even
better fit as it has more free parameters: $\chi^2$/DoF $
=158.75/180=0.88$ and $Q=87\%$ obtained for the values
$\Omega_{\rm m}=1-\Omega_\Lambda=0.34\pm0.04$. The best-fitting
model{ given in Equation} (\ref{eq:d_L}) and the $\Lambda$CDM
model have been compared with this {data }sample in {Figure~}1.
%11111111

\begin{figure}%[h!!]

\vs \centering
\resizebox{4in}{!}{\includegraphics{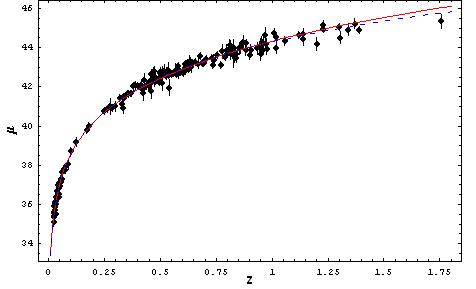}}
%\resizebox{3.1in}{!}{\includegraphics%{MS1272SNeIa
%{fig1.eps}}
%\includegraphics[width=9cm]{SNeIa.eps}

\vspace{-3mm} \caption{\baselineskip 3.6mm The `new gold sample'
of 182 SNeIa from Riess et al. (\citeyear{riess}) is compared with
some best-fitting models. The solid curve corresponds to {the
}model (\ref{eq:d_L}) and the dashed curve corresponds to the
spatially-flat $\Lambda$CDM model $\Omega_{\rm
m}=1-\Omega_\Lambda=0.34\pm0.04$. }
\end{figure}

\subsection{Observations of High-Redshift Radio Sources}

Let us now consider the data on the angular size and redshift of
radio sources compiled by Jackson \& Dodgson (\citeyear{radio}),
which %has
have 256  sources with their redshift in the range 0.5--3.8. These
sources are ultra-compact radio objects {with} angular sizes of
the order of a few milliarcseconds (mas), deeply embedded in
galactic nuclei and have a very short lifetime compared with the
age of the universe. Thus they are expected to be free from
evolutionary effects and hence may be treated as standard rods, at
least in the statistical sense. These sources are distributed into
16 redshift bins, {with }each bin containing 16 sources. This
compilation has recently been used by many authors to test
different cosmological models \citep{ban, vishwa_cqg,
vishwa-singh, vishwa_nuovo}. In order to fit the data to the
model, we derive the $\Theta-z$ relation in the following. The
angle $\Theta$ subtended {in a} telescope, by a source {with} the
proper diameter $d$, is given by \be
\Theta(z)=\frac{0.0688dh}{H_0 d_{\rm A}} %{\rm milliarcsecond}
~{\rm mas}, \ee where $d$ is measured in pc, $h$ is the present
value of the Hubble parameter in units of 100 km s$^{-1}$
Mpc$^{-1}$, and the angular diameter distance $d_{\rm A}=d_{\rm
L}/(1+z)^2$.

We find that the present model has a satisfactory fit to the data
with $\chi^2$/DoF $= 20.78/15=1.39$ and $Q = 14\%$. In order to
compare, we find that the best-fitting $\Lambda$CDM model has a
slightly better fit: $\chi^2$/DoF $= 16.03/14=1.15$ and $Q = 31\%$
obtained for the values $\Omega_{\rm
m}=1-\Omega_\Lambda=0.21\pm0.08$. These models are shown in
Figure~2. %222222222222

\begin{figure}
\vs \centering
\resizebox{4in}{!}{\includegraphics{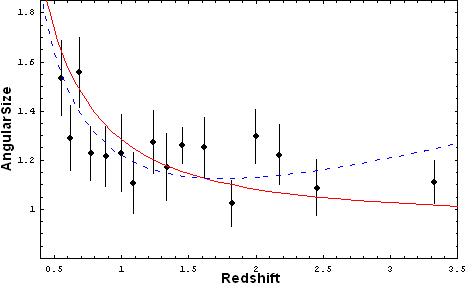}}
%\centering \resizebox{3.1in}{!}{\includegraphics{%MS1272
%AngS.eps}}
%\includegraphics[width=9cm]{AngS.eps}

\vspace{-3mm}

\caption{\baselineskip 3.6mm The data on the ultra-compact radio
sources compiled by Jackson \& Dodgson (\citeyear{radio}) are
compared with some best-fitting models. The solid curve corresponds
to the model represented by Equation~(\ref{eq:milne}) and the dashed
curve corresponds to the spatially-flat $\Lambda$CDM model
$\Omega_{\rm m}=1-\Omega_\Lambda=0.21\pm0.08$. }
\end{figure}

\subsection{Observations on $H_0$ and the Age of the Oldest Objects}

The age of the universe $t_0$, in big bang-like theories, is the
time{ that has} elapsed since the big bang. It depends on the expansion
dynamics of the model and is given by \be t_0=\int_0^\infty
\frac{H^{-1}(z)}{(1+z)}dz.\label{eq:age} \ee Hence, the Hubble
parameter controls the age of the universe, which in
{turn} depends on the free parameters of the model.
For example, in the standard cosmology, $H(z) =H_0\{\Omega_{\rm
m}(1+z)^3+\Omega_\Lambda+(1-\Omega_{\rm
m}-\Omega_\Lambda)(1+z)^2\}^{1/2}$. Although  $t_0$ is a model-based
parameter, a lower limit is put on it by requiring that the universe
must be at least as old as the oldest objects in it. This is done
through $t_{\rm GC}$, the age of globular clusters in the Milky
Way which are among the oldest objects we{ know} so far. The parameter
$H_0$ can be estimated in a model-independent way, for example, from
the observations of the low-redshift SNeIa, in which case the
predicted magnitude does not depend on the model-parameters. One can
use this value to calculate the age of the universe in a particular
theory which {can} be compared with the age of the oldest objects.
Thus the measurements of $H_0$ and $t_{\rm GC}$ provide a powerful
tool to test the underlying theory.

For example, by using the current measurements of $H_0 = 71\pm 6$ km
s$^{-1}$ Mpc$^{-1}$ from the Hubble Space Telescope Key Project
\citep{HST}, Equation~(\ref{eq:age}) gives $t_0$ for the
Einstein-de{ }Sitter model ($\Omega_{\rm m}=1$, $\Lambda=0$) as $9.18$
Gyr. This cannot be reconciled with the age of the oldest globular
cluster estimated to be  $t_{\rm GC}=12.5 \pm 1.2$ Gyr
\citep{Gnedin} and the age of the Milky Way as $12.5 \pm 3$ Gyr
coming from the latest uranium decay estimates \citep{Cayrel}.
However, for the concordance $\Lambda$CDM model with $\Omega_{\rm
m}=1-\Omega_\Lambda=0.27$ (as estimated by the WMAP project
\citep{wmap}), Equation~(\ref{eq:age}) gives a satisfactory age of
the universe{ as} $t_0=13.67$ Gyr which is well above the age of the
globular clusters.  The age of the universe in the present model is
given by $t_0=H_0^{-1}$, as can be checked from{ Equation} (\ref{eq:scale}).
For the above-mentioned value of $H_0$, this gives $t_0=13.77$ Gyr
which is even higher than the{ value from the} concordance model.

%\begin{table*}
\begin{table}
\centering

\begin{minipage}{100mm}\caption{Magnitudes, with
Uncertainties, of 13 HII Starburst
Galaxies\label{tabl}}\end{minipage}

\fns
\renewcommand\baselinestretch{1.6}

\fns\tabcolsep 5.5mm
\begin{tabular}{lcc}
\hline\noalign{\smallskip}
Galaxy & $z$ & $\mu \pm \sigma$ \\
\noalign{\smallskip}\hline\noalign{\smallskip}
Q0201-B13 & 2.17 & $47.49^{+2.10}_{-3.43}$ \\
Q1623-BX432 & 2.18 & $45.45^{+1.97}_{-3.07}$ \\
Q1623-MD107 & 2.54 & $44.82^{+0.31}_{-1.58}$ \\
Q1700-BX717 & 2.44 & $46.64^{+0.31}_{-1.58}$ \\
CDFa C1 & 3.11 & $45.77^{+0.31}_{-1.58}$ \\
Q0347-383 C5 & 3.23 & $47.12^{+0.44}_{-0.32}$ \\
B2 0902+343 C12 & 3.39 & $46.96^{+0.71}_{-0.81}$ \\
Q1422+231 D81 & 3.10 & $48.81^{+0.38}_{-0.40}$ \\
SSA22a-MD46 & 3.09 & $46.76^{+0.56}_{-0.51}$ \\
SSA22a-D3 & 3.07 & $49.71^{+0.43}_{-0.41}$ \\
DSF2237+116a C2 & 3.32 & $47.73^{+0.25}_{-0.25}$ \\
B2 0902+343 C6 & 3.09 & $45.22^{+1.38}_{-1.76}$ \\
MS1512-CB58 & 2.73 & $47.49^{+1.22}_{-1.57}$ \\
\noalign{\smallskip}\hline
\end{tabular}
\end{table}

\begin{figure}
\vs \centering
\resizebox{4in}{!}{\includegraphics{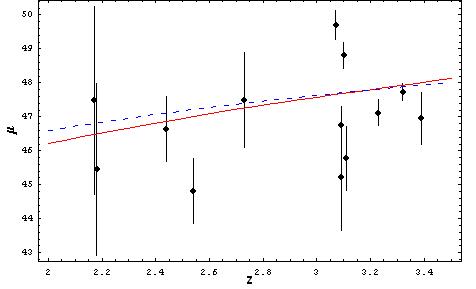}}
%\centering \resizebox{3.3in}{!}{\includegraphics{%MS1272gal
%fig3.eps}}
%\includegraphics[width=9cm]{gal.eps}

\vspace{-3mm} \caption{\baselineskip 3.6mm The HII-like starburst
galaxy data \citep{ratra} {are} compared with some best-fitting
models. The solid curve corresponds to the model represented by
Equation~(\ref{eq:d_L}) and the dashed curve corresponds to the
$\Lambda$CDM model{ with} $\Omega_{\rm m}=0.19${ and}
$\Omega_\Lambda=0.98$. }
\end{figure}

\subsection{Observations of Starburst Galaxies}

Let us now consider the data on the apparent magnitude and
redshift of starburst galaxies. Recent work has indicated that HII
starbur{s}t galaxies might be considered as  standard candles
because of a correlation between their velocity dispersion, H
luminosity and metallicity (see, for example, \cite{ratra} and
references therein). Siegel et al. (\citeyear{siegel}) compiled a
sample of 15 HII-like starburst galaxies with redshifts in the
range 2.17--3.39 (by using the data available in the literature)
in order to constrain $\Omega_{\rm m}$. Mania \& Ratra
(\citeyear{ratra}) modified this sample by excluding two HII
galaxies (Q1700-MD103 and SSA22a-MD41) that show signs of a
considerable rotational velocity component and used the resulting
sample (shown in Table~1) to constrain the cosmological models of
dark energy.

For this sample, the present model provides the minimum value of
$\chi^2$/DoF = 53.54/12 = 4.46 with $Q=3.4\times10^{-7}$, whereas
the standard $\Lambda$CDM model gives the minimum $\chi^2$/DoF =
53.32/10 = 5.33 with $Q=6.5\times10^{-8}$ for the values
$\Omega_{\rm m}=0.19${ and} $\Omega_\Lambda=0.98$. It would not be
fair to claim that any of these models fit the data well. Perhaps
the inherent scatter of the data is large, and a large sample size
is required to perform the test and to get any meaningful
constraint on the cosmological parameters. However, {it} is clear
from the fitting results that {compared to that of the standard
model, }the performance of the present model is better. The
results are shown in Figure~3. %3333333333333

\section{Evasion of the Problems of Standard Cosmology}

The field equation $R^{\mu\nu}=0$ registers success, not only on
the observational front, but also on the theoretical front.  As we
shall see in the following, the theory circumvents the
long-standing problems of standard cosmology, for example, the
horizon, flatness and the cosmological constant.

\subsection{Horizon Problem}

The (particle) horizon distance, given by \be d_{\rm H}(t)=S(t)
\int^t_0\frac{cdt'}{S(t')},\label{eq:hor} \ee sets a limit {on} the
observable or causally connected part of the universe at time
$t$. As a finite value exists for $d_{\rm H}$ in the standard
cosmology, this means that the universe has a horizon in this
theory. This is in conflict with the observed smoothness of the
cosmic microwave background (CMB) at the largest scales in all
directions, indicating that even the parts of the universe outside
the horizon have been in causal contact. Since no physical process
propagating at or below light speed could have brought them into
thermal equilibrium, it appears that the universe required special
initial conditions, which {are} supposed to be provided by inflation.
This problem does not exist in the present theory as $d_{\rm
H}=\infty$ at any time, as can be checked from{ Equations} (\ref{eq:scale}) and
(\ref{eq:hor}). Hence, the whole universe is always causally
connected, which explains the observed uniformity of CMB without
invoking the hypothetical inflaton field.

\subsection{Flatness and Cosmological Constant Problems}

The flatness problem of standard cosmology requires the initial
density of matter (represented by the energy-stress tensor) to be
extremely fine-tuned to its critical value (corresponding to a
spatially flat universe).  Even a tiny deviation from this value
would have had drastic effects on the nature of the present
universe. This problem is evaded in the present model owing to the
fact that the matter tensor does not explicitly{ appear} in the
dynamical equations and all the fields are represented through the
geometry.

 The cosmological constant problem is circumvented for the same reason because
  {its} origin lies in a conflict between the energy-stress
tensor and the vacuum expectation values derived from quantum
field theory. Let us recall that the cosmological constant is
represented by the energy-stress tensor of a perfect fluid (through
a particular equation of state) and hence it is absent in the
present theory. Hence, any other candidate of dark
energy also{ does not exist} in the present theory.

\section{Gravity of the `vacuum' field equation $R^{\mu\nu}=0$}

It thus seems that equation $R^{\mu\nu}=0$ get{s} strong support
not only from the observations but also from the theory. One may
ignore this  as chance happenings. However, we adopt the view that
it is unlikely to have so many coincidences happen together and
the theoretical as well as observational {evidence }supports
equation $R^{\mu\nu}=0$ at all scales{; these} perhaps point
towards some missing link of the theory, unnoticed so far. Let us
then see if equation $R^{\mu\nu}=0$ can describe the real universe
with matter, as the observations suggest. As the source of
curvature is invariably matter, %-energy,
 there have already been
evidences available since the very inception of the theory, from a
variety of solutions of equation $R^{\mu\nu}=0$, which indicate
that the space described by these equations {does not necessarily
}need {to} be empty. For example, let us consider the following
well-known solutions of equation $R^{\mu\nu}=0$ which have
non-vanishing curvature.

\subsection{Curved Solutions of $R^{\mu\nu}=0$}

\noindent
{\bf Schwarzschild Solution:}

\noindent Discovered by Karl Schwarzschild in 1915 immediately after
GR was formulated, the solution forms the cornerstone of GR. It is
believed to represent the spacetime structure outside an isotropic
mass in an empty space \citep{hawking-ellis} \be
 ds^2=\left(1+\frac{K}{r}\right)c^2 dt^2-\frac{dr^2}{(1+K/r)}-r^2d\theta^2-r^2\sin^2\theta ~d\phi^2,\label{eq:Sch}
\ee where $K$ is a constant of integration. In fact, all the
experiments which have so far been carried out to test GR are based
on the predictions by this solution (except the Gravity Probe B
experiments, which are based on the predictions of the Kerr
solution).

\medskip
\noindent
{\bf Kerr Solution:}

\noindent
Discovered by Roy Kerr in 1963, the solution describes the spacetime surrounding a spherical mass $m$  spinning with angular momentum per unit mass = $\alpha$ (so that its total angular momentum $=mc\alpha$). In the Boyer-Lindquist coordinates  \citep{hawking-ellis}, the solution takes the form
\[
ds^2=\left(1-\frac{r_{\rm S}r}{\rho^2}\right)c^2 dt^2-\frac{\rho^2}{\Delta}dr^2-\rho^2d\theta^2 -\left(r^2+\alpha^2+\frac{r_{\rm S}r\alpha^2}{\rho^2}\sin^2\theta\right)\sin^2\theta ~d\phi^2
\]
\be
+\frac{2r_{\rm S}r\alpha}{\rho^2}\sin^2\theta ~d\phi ~cdt,~~~~~~~~~~~~~~~~~~~~~~~~~~~~~~~~~~~~~~~~~~~~~~~~~~~~~~~~~~~~~~~~~~~~~~~~\label{eq:kerr}
\ee
where $\rho^2=r^2+\alpha^2 \cos^2\theta$, $\Delta=r^2-r_{\rm S}r+\alpha^2$ and $r_{\rm S}=2Gm/c^2$ is the Schwarzschild radius. When $\alpha=0$, the solution reduces to the Schwarzschild solution.

It may be mentioned that solutions (\ref{eq:Sch}) and
(\ref{eq:kerr}) cannot be transformed to the flat Minkowski metric
by any possible coordinate transformation, as the Riemann-Chistoffel
curvature tensor $R_{\lambda\mu\nu\delta}\ne0$ in these cases. Let
us decipher the source of curvature in these solutions.

\subsection{Sources of Curvature in  $R^{\mu\nu}=0$}

As Equation~(\ref{eq:RicciEq}) is devoid of the source term
$T^{\mu\nu}$, one may wonder from where {the }solution
(\ref{eq:Sch}) derives its curvature. It is believed that the
mystery of the presence of this curvature is related to the central
singularity of the spherically symmetric space represented by
(\ref{eq:Sch}) (which is associated, through a correspondence
between the Newtonian and Einsteinian theories of gravitation in the
case of a weak field, {with} an isotropic mass sitting at the
cente{r} $r=0$). Thus, the source of curvature in Einstein's theory
is regarded to be either $T^{\mu\nu}$ or a singularity in
$g_{\mu\nu}$. However, it should be noted that{ the} metric
(\ref{eq:Sch}) represents space exterior to the central mass at
$r=0$ and {\it not} the point $r=0$ itself, where the metric breaks
down. So, how can a mass situated at the point $r=0$ (which is not
even represented by the metric) curve the space of (\ref{eq:Sch}) at
points for which $r>0$? Obviously, one cannot expect the Newtonian
theory of action-at-a-distance to work in the framework of GR which
is a local theory.

A little reflection suggests that the agent responsible for the
curvature in (\ref{eq:Sch}) at the points for $r>0$ must be the
gravitational energy, which can definitely exist in an empty space.
However, if this is true, one should be able to calculate the
gravitational energy from {the }metric (\ref{eq:Sch}). We can certainly
do this by the following two simple observations:
%\begin{quote}
\begin{enumerate}
\item[(1)] The metric (\ref{eq:Sch}) departs from flat spacetime
in the term $K/r$, implying that this term must be the source of
curvature.

\item[(2)] We have shown that the source of curvature in
(\ref{eq:Sch}), at the points $r>0$, must be the gravitational
energy.
\end{enumerate}
%\end{quote}

\noindent Taken together, these two points imply that $K/r$ must
be the gravitational energy (in the units with $c=1$) in
(\ref{eq:Sch}). This is in perfect agreement with the way the
value of the constant $K$ is determined. Let us recall that the
constant $K$ in Equation~(\ref{eq:Sch}) is specified in terms of
the Newtonian gravitational potential energy (by requiring that in
the case of a weak gravitational field, Newton's law should hold)
giving $g_{00}=1+2\psi/c^2$ where $\psi=-Gm/r$ is the
gravitational energy (per unit mass) at a distance $r$ from the
central mass $m$ producing the field. It is thus established that
the source of the curvature of spacetime in (\ref{eq:Sch}) is the
energy of the gravitational field present at the points exterior
to $r=0$.

However, the fact remains %there
 that no formulation of the
gravitational energy has been incorporated in equation
$R^{\mu\nu}=0$ ({or} in Eq.~(\ref{eq:EinsteinEq})). This implies
that the gravitational energy already exists there implicitly in the
geometry, through the non-linearity of the field equations, and no
additional incorporation thereof is needed. This fits very well in
the story of the failure to discover an energy-stress tensor for the
gravitational field\footnote{It may be mentioned that despite the
century-long dedicated efforts of many luminaries, the attempts to
discover a unanimous formulation of the gravitational field energy
in GR {have} failed. {This is p}rimarily because of the %the
 non-tensorial
character of the energy-stress pseudotensors of the gravitational
field and the lack of a unique agreed-upon formula for it. Secondly,
{this is }because of the inherent difficulty in the localization of the
gravitational energy.}. As mentioned earlier, the
energy-stress tensor of the gravitational field is not included in
$T^{\mu\nu}$ in Einstein's field Equation~(\ref{eq:EinsteinEq}), as
this tensor does not exist. Here we find the answer to this mystery
- the energy-stress tensor of the gravitational field does not exist
simply because the tensor is not needed in the geometric framework
of GR{;} it already exists there inherently in the geometry of
equation $R^{\mu\nu}=0${.} We should note that in the
weak-field approximation, the GR equations do reduce to the usual
Newtonian dynamical equations; gravitational energy, force, etc. do
emerge respectively from the metric tensor, the Christoffel symbol,
etc., {\it without adding any formulation of the gravitational field
energy to the Einstein equations}.

The {insight gained }above about the futility of the energy-stress
tensor of the gravitational field is also corroborated by the Kerr
metric{ given in Equation} (\ref{eq:kerr}) wherein the angular momentum also contributes
to its curvature. This is an entirely new
{occurrence} for GR. It may be mentioned that
there is no place for the angular momentum in $T^{\mu\nu}$ in the
framework of Einstein's theory, which needs to be extended to
non-Riemannian curved spacetime with torsion (as in the
Einstein-Cartan theory) to support asymmetric Ricci and metric
tensors, so that an asymmetric energy-stress tensor of spin can
appear on the right hand side of the field equations. (However, when
the right hand side is vanishing, the Ricci tensor need not be
asymmetric and the Einstein-Cartan `vacuum' equations reduce to
$R^{\mu\nu}=0$.)

One may not show much inhibition to agree that{ expressions given by}
Equation~(\ref{eq:RicciEq}) do contain energy of the gravitational
field (by leverag{ing} the controversial nature of the subject{,}
as has been{ done}), despite the monumental works of Tolman, Papapetrou,
Landau-Lifshitz, M${\o}$ller and Weinberg on the (pseudo)
energy-stress tensors of the gravitational field. However, one would
maintain that although{ expressions defined by} Equation~(\ref{eq:RicciEq}) contain energy
of the gravitational field, they describe spacetime structures in an
{\it otherwise empty} space.  Further, the source of curvature in
the solutions of Equation~(\ref{eq:RicciEq}) must be the
singularity which fuels the gravitational energy. Let us now
consider another curved solution of Equation~(\ref{eq:RicciEq}),
which may not fit this interpretation{ very well}.

\medskip
\noindent
{\bf Kasner Solution:}

\noindent This important cosmological solution of equation
$R^{\mu\nu}=0$ was discovered by Edward Kasner in
\citeyear{kasner}. Later it was rediscovered by  V. V.
Narlikar and K. R. Karmarkar in \citeyear{curious} and
again by Abraham Taub in \citeyear{taub}. The
solution, which is a curved Bianchi type I metric, describes a model
universe which is homogeneous but anisotropic. The solution is given
by \be ds^2=c^2 dt^2- t^{2p_1}dx^2- t^{2p_2}dy^2-
t^{2p_3}dz^2,\label{eq:kasner} \ee where the constants $p_1$, $p_2$
and $p_3$ satisfy
\[
p_1+p_2+p_3=1, ~~p_1p_2+p_2p_3+p_3p_1=0.
\]
It should be noted that the metric admits a
singularity\footnote{The form of the Kasner metric discovered by
\cite*{curious} is \be ds^2=c^2 dt^2- (1+nt)^{2p_1}dx^2-
(1+nt)^{2p_2}dy^2- (1+nt)^{2p_3}dz^2,\label{eq:curious} \ee where
$n$ is a constant. We note that by the use of the transformations
$nt=1+n\bar{t}, ~x= n^{p_1}\bar{x}, ~y= n^{p_2}\bar{y}$ and
$z=n^{p_3}\bar{z}$, the metric (\ref{eq:kasner}) takes the form
(\ref{eq:curious}) in the new coordinates. As the singularity in
(\ref{eq:curious}) now appears at $t=-1/n$, it seems that it can
be avoided with a positive $n$. However, as the average scale
factor (= spatial volume$^{1/3}=1+nt$) has the same behavior in
the interval $-1/n\leq t \leq 0$ as it has for $0\leq t$ (i.e., no
bouncing behavior at $t=0$), this feature to push the singularity
back before the time $t=0$ just appears {as} a rescaling of time.}
at $t=0$ (note that all of the exponents $p_1$, $p_2$ and $p_3$
are not positive).

The usual  interpretation provided to (\ref{eq:kasner}) is that it
represents an empty homogeneous universe in which the space is
expanding and contracting (anisotropically) at different rates in
different directions (for example, for $p_1=p_2=2/3$ and $p_3=-1/3$,
the space is expanding in two directions and contracting in the
third). As we have noted, the solution (\ref{eq:kasner}) contains a
singularity at $t=0$ but not at any other time. However, a past
singularity, which does not exist now, fueling the gravitational
energy  now without any other source, does not seem compatible with
the understanding of the gravitational energy. Further, the
curvature in (\ref{eq:kasner}) is expected to have contributions
from the net non-zero momentum resulting from the anisotropic
expansion/contraction of the homogeneous space. However, it does not
make much sense to imagine momentum resulting from the
expanding/contracting {\it empty} space with no matter. It does not
make sense, in the first place, to think of expanding/contracting
`homogeneous' space {\it without} matter.

As has been shown above, {expressions from }equation $R^{\mu\nu}=0$
do reveal gravitational energy, through geometry, in its curved
solutions (\ref{eq:Sch}) and (\ref{eq:kerr}). Hence, the third
curved solution  (\ref{eq:kasner}) is also expected to contain the
gravitational energy. However, unlike the solutions{ expressed in
Equations} (\ref{eq:Sch}) and (\ref{eq:kerr}) (which represent space
outside the source mass), the cosmological solution
(\ref{eq:kasner}) cannot be expected to have any `outside{.}' Since
the ultimate source of the gravitational field is matter, this
implies that the source matter fields (together with the resulting
gravitational field) must also be {inherently }contained in solution
(\ref{eq:kasner}){.} Does it then mean that, like the gravitational
field, the matter field is also inherently{ present} in equation
$R^{\mu\nu}=0$? Let us postpone this question until its answer
emerges as a natural consequence from the following issue.

\subsection{On the Flatness of Solution (\ref{eq:milne})}

It would be natural to ask why solution (\ref{eq:milne}) is flat
while the other solutions  (\ref{eq:Sch})$-$(\ref{eq:kasner}), of
the same equation $R^{\mu\nu}=0$, are curved. [It should be noted
that, by the use of the transformations $\bar{t}=t\sqrt{1+r^2}$ and
$\bar{r}=ctr$,  metric (\ref{eq:milne}) can be brought to manifestly
Minkowskian form in the coordinates $\bar{t},\bar{r}, \theta, \phi$
\citep{narlikar}.]

One may argue that solution (\ref{eq:milne}) is Minkowskian simply
because $T^{\mu\nu}$ is zero in (\ref{eq:RicciEq}). However, if
this is so, why do we get curved solutions
(\ref{eq:Sch})$-$(\ref{eq:kasner}) from the same equation
$R^{\mu\nu}=0$? If {expressions from }equation  $R^{\mu\nu}=0$
contain {a }gravitational field{ that acts as the} source of
curvature present in solutions (\ref{eq:Sch})$-$(\ref{eq:kasner}),
they must also do so in solution (\ref{eq:milne})%also
\footnote{Equations{ from} $R^{\mu\nu}=0$ are not competent
enough to decipher the source term $m$ in the Schwarzschild and
Kerr solutions  (\ref{eq:Sch}) and (\ref{eq:kerr}), just from
their symmetries. Rather, this is done through an additional
constraint that GR should reduce to Newtonian gravitation in the
case of a weak stationary gravitational field, as has been
mentioned earlier. Hence, taken on face value, the other two
solutions of $R^{\mu\nu}=0$, viz. (\ref{eq:milne}) and
(\ref{eq:kasner}), must also have the same status and it is also
quite probable to assign fields to them, in a manner consistent
with their symmetries.}.

Obviously, solutions (\ref{eq:Sch})$-$(\ref{eq:kasner}) have
singularities to fuel the gravitational field, %while
but solution (\ref{eq:milne}) does not. {However}, what is there
to stop the singularity {from} occur{ring} in (\ref{eq:milne})?
The only difference between the solutions  (\ref{eq:milne}) and
(\ref{eq:Sch})$-$(\ref{eq:kasner}) is that they have different
types of symmetries in their spacetime structures. While metric
(\ref{eq:milne}) is homogeneous and isotropic, metrics
(\ref{eq:Sch})$-$(\ref{eq:kasner}) are either inhomogeneous or/and
anisotropic. However, how a relaxation in the homogeneity and
isotropy can result in a singularity cannot be answered by
conventional wisdom. A possible explanation to the present
situation leads us to the following two possibilities.

\begin{enumerate}

\item[(1)] Equations{ arising from} $R^{\mu\nu}=0$ represent {an
}empty spacetime structure and can support curved as well as flat
spacetime solutions. However, they are unable to explain how a
solution acquires curvature or flatness. For instance, solutions
(\ref{eq:kasner}) and (\ref{eq:milne}) represent similar spacetime
structures with the only difference{ being} that while the
homogeneous space in (\ref{eq:kasner}) is expanding and
contracting in different directions at different rates, the same
space is expanding or contracting isotropically in
(\ref{eq:milne}). How this difference accounts for their curved
and flat states{,} and controls the appearance of the singularity,
cannot be explained by equations{ from} $R^{\mu\nu}=0$.

\item[(2)] The geometry of equation $R^{\mu\nu}=0$ does contain
impressions of the gravitational as well as the matter fields. The
structure of the geometry, of a chosen matter distribution, is
determined by the net contribution from the material and the
gravitational fields which, if non-zero, may be manifested in the
guise of a singularity (which may be considered a
general-relativistic analog of the `source').

\end{enumerate}

Though the second possibility appears baffling and orthogonal to
usual understanding, it provides not only a sensible meaning to the
Kasner solution  (\ref{eq:kasner}) in the absence of the
energy-stress tensors of the gravitational or the matter fields, but
also a reasonable explanation to the flatness of solution
(\ref{eq:milne}), as we see in the following. If we believe that
Equation~(\ref{eq:RicciEq}) inherently{ contains} material and
gravitational fields, solution (\ref{eq:milne}) would then represent
homogeneously distributed matter throughout the space at all times.
As the positive energy of the matter field would be exactly
balanced, point by point,  by the negative energy of the resulting
gravitational field  (contrary to the case of the Schwarzschild
solution where there is only the gravitational energy and no matter
at the points represented by the metric), %{which}
this would provide a net vanishing energy. {T}here would{ also
not} be any momentum contribution from the isotropic
expansion/contraction of the material system (contrary to the case
of the Kasner solution). Hence, in the absence of any net non-zero
energy, momentum or angular momentum, the spacetime of
(\ref{eq:milne}) would not have any curvature.

This implies that it is the symmetry of the chosen spacetime
structure which determines whether a solution of $R^{\mu\nu}=0$
will be curved (may have a singularity) or flat (without{ a}
singularity). This is in perfect agreement with the appearance of
different kinds of singularities, in accordance with the chosen
symmetries in the solutions: while the Schwarzschild solution
(describing the spacetime structure exterior to a point mass) has
a point singularity, the Kerr solution (describing the spacetime
structure exterior to a rotating mass) has a ring singularity{;}
the Kasner solution (in which the $t$ = constant hypersurfaces are
expanding and contracting at different rates in different
directions) presents an oscillating kind of singularity %of
  at $t=0$.

The discovery of the net vanishing energy-momentum-angular
momentum
 in a homogeneous distribution of matter expanding/contracting
isotropically, appears {to be }consistent with several
investigations and results which indicate that the total energy of
the universe is zero. Thus, a flat spacetime solution, which has so
far been a notion of special relativity (SR), can be achieved in the
real universe in the presence of matter, which {dynamically
}originates from the field equations, and is not assumed a priori
(or just %put
added by hand) as in SR. This new result also seems consistent
with the theories of inflation which predict a nearly flat
universe after expanding it by a factor of $10^{78}$ in just
$10^{-36}$ seconds, leaving a nearly flat spacetime. Further, the
appearance of a flat spacetime in the presence of matter is also
not impossible in the conventional approach. For example, it has
been shown by \cite{ayon} that conformally coupled matter does not
always curve spacetime.

It may be mentioned that solution (\ref{eq:milne}) is generally
considered as the Milne model, which is not quite correct when taken
in the traditional context of the empty universe. Although the
evolution{ary} dynamics of the Milne universe (based on Milne's
kinematical relativity with foundations different from those of GR)
{are} the same as {those} given by the empty Friedman model, it is
not empty. Rather the Milne model assumes the Minkowskian spacetime
filled with matter wherein the matter does not interact with the
geometry due to some unknown reasons. The present theory, governed
by the field equation $R^{\mu\nu}=0$, not only explains why the
homogeneous, isotropic universe is Minkowskian, but also predicts
when a curved solution is also possible.

Thus the conclusion is that the spacetime already contains the
`field{,}' which must not be inserted again in the field
equations of gravitation through the formulation of, for example,
$T^{\mu\nu}$. It should be noted that being a geometric theory of
gravitation, GR eliminates any possibility to represent gravitation
in terms of a force. Rather the theory replaces the effects of the
force through geometry. Similarly, the effects of stresses,
momenta, angular momenta and energy are {also
}revealed through geometry. Though this surprising
discovery may appear too revolutionary to digest, {it }is in
striking agreement with and corroborated by the recent studies on
the relativistic formulation of matter which indicate that, like the
energy-stress tensor of the gravitational field, a flawless
energy-stress tensor of the matter fields {also does not} exist. This
issue {is} discussed in brief in the next section.

\subsection{On the Formulation of Matter by $T^{\mu\nu}$}

We have seen above that the appearance of the gravitational energy
through the geometry of equation $R^{\mu\nu}=0$, without adding any
formulation of the energy-stress tensor of the gravitational field
to this equation, is vindicated by the absence of a proper
energy-stress tensor of the gravitational field. Do we have any
similar evidence from the energy-stress tensor of the matter fields?
Yes{ we do.} Recently, it has been shown
\citep{vishwaApSS2} that a critical analysis of the formulation of
matter given by $T^{\mu\nu}$ reveals some surprising inconsistencies
and paradoxes. Corrections have been discovered to rectify the
problems, which however render the theory incompatible with many
observations \citep{vishwaApSS2}. This implies that the relativistic
formulation of matter fields given by the energy-stress tensor
$T^{\mu\nu}$ is not consistent with the geometric description
of GR. In fact, the relativistic formulation of matter in terms of
$T^{\mu\nu}$ has some more fundamental inconsistencies (besides
those discovered in \citealt{vishwaApSS2}) which have not been
realized earlier. We mention the following two.

\begin{enumerate}
\item[(1)] The general expression of the energy-stress tensor
$T^{\mu\nu}$ is obtained by deriving it first in the absence of
gravity, i.e., in SR. It is then imported to the actual case in
the presence of gravity, by making use of an inertial observer,
which exists admittedly at all points of spacetime (by courtesy of
the principle of equivalence). Formulating a tensor representation
of the fluid element in a small neighborhood of the observer, the
expression of the tensor in the presence of gravity is imported,
from SR to GR, through a coordinate transformation. This is the
standard way to derive $T^{\mu\nu}$ \citep{vishwaApSS2}. However,
the simple point which has not been noticed, which makes this
derivation questionable, is that an inertial coordinate system is
valid only at a point, and {\it not} in a neighborhood, however
small it is (Christoffel symbols can be made vanishing only at a
point, and {\it not} in a neighborhood). {However, because} a
fluid element cannot be defined at a point, we do need a
neighborhood. Hence, at best, the tensor $T^{\mu\nu}$ may
approximate a dust (with vanishing pressure) but cannot represent
a fluid (with non-zero pressure) which requires more than one
particle to generate pressure.

\item[(2)] As the derivation of $T^{\mu\nu}$ assumes its validity
in the absence of gravitation (in a flat spacetime), this
{becomes} contradictory to the very notion of $T^{\mu\nu}$ being
the source of curvature. To exemplify this, let us note that in a
flat spacetime, the left hand side of
Equation~(\ref{eq:EinsteinEq}) vanishes automatically, but not the
right hand side which has to be %put
made equal to zero by hand. That is, the source of curvature can
exist there without producing any curvature{.} Although
Equation~(\ref{eq:EinsteinEq}), being a geometric formulation of
gravitation, is expected to be valid in a curved spacetime, it
must also {consistently }reduce to the no-gravitation case. This
provides another reason why $T^{\mu\nu}$ should not appear in the
field equations of gravitation.

\end{enumerate}

It should be noted that the relativistic description of the matter
given by the energy-stress tensor $T^{\mu\nu}$ has never been tested
in any direct experiment.  As has been mentioned earlier, the
classical tests of GR consider $T^{\mu\nu}=0$ and test only the
geometric aspect of GR given by equation $R^{\mu\nu}=0$. The
above-mentioned theoretical crisis in the relativistic formulation
of matter acquires a new meaning in the present context and implies
that the concept of the energy-stress tensor (be it of{ a} matter
field or{ a} gravitational field) is not compatible with the
geometric formulation of gravitation, simply because it is already{
inherently} included in the geometry{.}

A similar view was expressed about four decades ago by J. L. Synge,
one of the most distinguished mathematical physicists of the 20th
Century: {\it ``The concept of energy-momentum tensor is simply
incompatible with general relativity{.}"}

\section{Discussion and Conclusions}

In spite of various observational verifications of GR, deep
mysteries continue to haunt our theoretical understanding of the
ingredients of the energy-stress tensor in the form of
{the dark sectors$-$inflaton}, dark matter and dark
energy, which do not have any non-gravitational or laboratory
evidence and remain unidentified.

Motivated by this fact and guided by strong observational support
{for} the so called `vacuum' field equations{ expressed as}
$R^{\mu\nu}=0$, we develop an understanding that{ the} energy-stress
tensor is perhaps a redundant part of the Einstein field equations
and the source of gravitation is the geometry itself. A critical
analysis of the different solutions of equation $R^{\mu\nu}=0$
supports this view indicating that equation $R^{\mu\nu}=0$ {\it does
not} represent an empty spacetime and impressions of the
gravitational as well as material fields are inherently present in
the equations (though they do not play a direct role in the
dynamics), which are revealed through the geometry. Einstein
believed that {\it ``On the basis of the general theory of
relativity, space as opposed to `what fills space', has no separate
existence''} (\citealt{einstein}). If this is true, considering a
spacetime structure (conditioned by the equation $R^{\mu\nu}=0$)
must be equivalent to considering the accompanying fields (material
and gravitational) as well, and there should be no need to add any
extra formulation thereof to the field equations.

The fact that the sources of gravitation are implicitly present in
equations{ defined by} $R^{\mu\nu}=0$ and must not be added again is vindicated
by the failure to obtain a proper energy-stress tensor of the
gravitational field. It is further supported by a number of
paradoxes and inconsistencies discovered recently in the
relativistic formulation of matter given by the energy-stress tensor
$T^{\mu\nu}$ implying that, as in the case of the gravitational
field, a flawless proper energy-stress tensor of the matter fields
{also does not} exist \citep{vishwaApSS2}. This, in fact, leaves equation
$R^{\mu\nu}=0$ as the only possibility for a consistent field
equation of gravitation in the existing framework of GR. One may
argue that a consistent field equation of gravitation is expected to
reduce to Poisson's equation $\nabla^2\psi=4\pi G\rho$ in a weak
stationary gravitational field. However, this requirement, which is
already compromised in the concordance $\Lambda$CDM cosmology, no
longer seems obligatory. It should be noted that the Einstein field
equations with a non-zero\footnote{It would not be correct to claim
that $\Lambda$ is negligible in the concordance cosmology, in order
to give any advantage to this theory. In fact, the mass density
associated with $\Lambda$, say, $\rho_\Lambda=\Lambda c^2/8\pi G$,
is comparable with the matter density.} $\Lambda$ do {\it not }
fulfill this requirement \citep{weinberg}.

Although this entirely new insight about the geometry serving as the
source of gravitation in the metric theories of gravity may
appear orthogonal to the usual understanding, it is not only in
striking agreement with the theory and observations, but also
provides natural explanations to some unexplained puzzles.
Additionally, it removes the long-standing problems of standard
cosmology, viz., the horizon, the flatness and the cosmological
constant.

It was Einstein's obsession that the vibrant geometrical part of GR
is `marble' and matter is `wood{,}' and that all attempts should be
directed to turn wood into marble. It finally turns out that
the `wood' is a redundant part of the theory whose departure
enhances the beauty of the `marble'  in the true field equation of
gravitation $R^{\mu\nu}=0$ due to its extreme
simplicity{.}

It may be mentioned that any proposed theory of gravitation,
supplying a model of the universe, is expected to explain the
observations of the CMB radiation and the baryon acoustic
oscillations (BAO). Though a detailed discussion on this subject
would require further study, it may be mentioned for the time being
that{,} tak{en} at face value, the only
unanimous prediction of the CMB observations is a flat spatial
geometry \citep{vishwa_mnras, cmb, wmap}. As has been mentioned
earlier, Equation~(\ref{eq:milne}) can be transformed to the
Minkowskian form, by using suitable transformations, which does have
a flat spatial geometry.

Additionally, we have shown that the universe is not empty in the
present theory, though the matter fields do not play a direct role,
hence providing full leverage {for} the parameters $\Omega_{\rm m}$,
$\Omega_{\rm b}$, etc., to fit the observations of CMB and BAO which
{indicate} that $\Omega_{\rm m}\approx 0.3$ \citep{komatsu, blake}.

Finally, it would be worthwhile to mention an interesting
interpretation (which is due to an anonymous referee) of the present
findings. In terms of the standard picture, the gravitating sources
can be viewed as a discrete distribution of point particles seen at
a microscopic level (in a restricted sense). Although this
interpretation would not hold in the case of the scalar fields or
the cosmological constant, it would not create any real problem. The
only obligatory scalar field required by the standard paradigm, to
explain the horizon and flatness problems, is the inflaton field,
which is {however} not required by the new theory, as we have seen
earlier. While the horizon problem does not exist in the new theory
as the whole universe is always causally connected, the flatness
problem is averted due to the absence of the matter tensor. The
cosmological constant or any other candidate of dark energy is
absent in the new theory, due to the same reason, as has been
explained earlier. Thus the alternative interpretation appears
promising {and }worth exploring further.

\begin{acknowledgements}
The author benefited {from} valuable
discussions with Professor Abdus Sattar and Professor Jayant
Narlikar. Thanks are due to Vijay Rai for sending some important
literature.
\end{acknowledgements}


\begin{thebibliography}{99}
\small \setlength{\itemindent}{-3mm} \setlength{\itemsep}{-0.5mm}
\setlength{\baselineskip}{4.7mm}

\bibitem[{{Ay{\'o}n-Beato} et~al.(2005){Ay{\'o}n-Beato}, {Mart{\'{\i}}nez},
  {Troncoso}, \& {Zanelli}}]{ayon}
{Ay{\'o}n-Beato}, E., {Mart{\'{\i}}nez}, C., {Troncoso}, R., \& {Zanelli}, J.
  2005, \prd, 71, 104037

\bibitem[{{Banerjee} \& {Narlikar}(1999)}]{ban}
{Banerjee}, S.~K., \& {Narlikar}, J.~V. 1999, \mnras, 307, 73

\bibitem[{{Blake} et~al.(2011){Blake}, {Kazin}, {Beutler} et~al.}]{blake}
{Blake}, C., {Kazin}, E.~A., {Beutler}, F., et~al. 2011, \mnras,
418, 1707

\bibitem[{{Blanchard}(2005)}]{cmb}
{Blanchard}, A. 2005,  Dark
{M}atter in {A}stro- and {P}article
{P}hysics, 34

\bibitem[{{Cayrel} et~al.(2001){Cayrel}, {Hill}, {Beers} et~al.}]{Cayrel}
{Cayrel}, R., {Hill}, V., {Beers}, T.~C., et~al. 2001, \nat, 409, 691

\bibitem[{Einstein(1920)}]{einstein}
Einstein, A. {1920}, Relativity: The
{S}pecial and the {G}eneral
{T}heory

\bibitem[{{Gnedin} et~al.(2001){Gnedin}, {Lahav}, \& {Rees}}]{Gnedin}
{Gnedin}, O.~Y., {Lahav}, O., \& {Rees}, M.~J. 2001, astro-ph/0108034

\bibitem[{{Hawking} \& {Ellis}(1973)}]{hawking-ellis}
{Hawking}, S.~W., \& {Ellis}, G.~F.~R. 1973, {The
{L}arge-scale {S}tructure of
  {S}pace-time} (Cambridge: Cambridge Univ. Press)

\bibitem[{{Jackson} \& {Dodgson}(1997)}]{radio}
{Jackson}, J.~C., \& {Dodgson}, M. 1997, \mnras, 285, 806

\bibitem[{Kasner(1921)}]{kasner}
Kasner, E. 1921, American Journal of Mathematics, 43, 217

\bibitem[{{Komatsu} et~al.(2011){Komatsu}, {Smith}, {Dunkley} et~al.}]{komatsu}
{Komatsu}, E., {Smith}, K.~M., {Dunkley}, J., et~al. 2011, \apjs, 192, 18

\bibitem[{{Larson} et~al.(2011){Larson}, {Dunkley}, {Hinshaw} et~al.}]{wmap}
{Larson}, D., {Dunkley}, J., {Hinshaw}, G., et~al. 2011, \apjs, 192, 16

\bibitem[{{Mania} \& {Ratra}(2012)}]{ratra}
{Mania}, D., \& {Ratra}, B. 2012, Physics Letters B, 715, 9

\bibitem[{{Mould} et~al.(2000){Mould}, {Huchra}, {Freedman} et~al.}]{HST}
{Mould}, J.~R., {Huchra}, J.~P., {Freedman}, W.~L., et~al. 2000, \apj, 529, 786

\bibitem[{{Narlikar}(2002)}]{narlikar}
{Narlikar}, J.~V. 2002, {An {I}ntroduction to {C}osmology}
{(}Cambridge: Cambridge Univ. Press{)}, 140

\bibitem[{Narlikar \& Karamarkar(1946)}]{curious}
Narlikar, V.~V., \& Karamarkar, K.~R. 1946, Current Science, {15}, 69

\bibitem[{{Perlmutter} et~al.(1999){Perlmutter}, {Aldering}, {Goldhaber}
  et~al.}]{perlmutter}
{Perlmutter}, S., {Aldering}, G., {Goldhaber}, G., et~al. 1999, \apj, 517, 565

\bibitem[{{Riess} et~al.(2007){Riess}, {Strolger}, {Casertano} et~al.}]{riess}
{Riess}, A.~G., {Strolger}, L.-G., {Casertano}, S., et~al. 2007, \apj, 659, 98

\bibitem[{{Siegel} et~al.(2005){Siegel}, {Guzm{\'a}n}, {Gallego}, {Ordu{\~n}a
  L{\'o}pez}, \& {Rodr{\'{\i}}guez Hidalgo}}]{siegel}
{Siegel}, E.~R., {Guzm{\'a}n}, R., {Gallego}, J.~P., {Ordu{\~n}a L{\'o}pez},
  M., \& {Rodr{\'{\i}}guez Hidalgo}, P. 2005, \mnras, 356, 1117

\bibitem[{{Taub}(1951)}]{taub}
{Taub}, A.~H. 1951, Annals of Mathematics, 53, 472

\bibitem[{{Vishwakarma}(2000)}]{vishwa_cqg}
{Vishwakarma}, R.~G. 2000, Classical and Quantum Gravity, 17, 3833

\bibitem[{{Vishwakarma}(2003)}]{vishwa_mnras}
{Vishwakarma}, R.~G. 2003, \mnras, 345, 545

\bibitem[{{Vishwakarma} \& {Singh}(2003)}]{vishwa-singh}
{Vishwakarma}, R.~G., \& {Singh}, P. 2003, Classical and Quantum Gravity, 20,
  2033

\bibitem[{{Vishwakarma}(2007)}]{vishwa_nuovo}
{Vishwakarma}, R.~G. 2007, Nuovo Cimento B Serie, 122, 113

\bibitem[{{Vishwakarma} \& {Narlikar}(2010)}]{critique}
{Vishwakarma}, R.~G., \& {Narlikar}, J.~V. 2010, \raa, 10, 1195

\bibitem[{{Vishwakarma}(2012)}]{vishwaApSS2}
{Vishwakarma}, R.~G. 2012, \apss, 340, 373

\bibitem[{{Weinberg}(1972)}]{weinberg}
{Weinberg}, S. 1972, {Gravitation and Cosmology: Principles and
Applications of
  the General Theory of Relativity} {(}John Wiley \& Sons{)}, 155

\end{thebibliography}
\end{document}